\documentclass[conference]{IEEEtran}
\IEEEoverridecommandlockouts

\usepackage{cite}
\usepackage{amsmath,amssymb,amsfonts}
\usepackage{algorithmic}
\usepackage{graphicx}
\usepackage{textcomp}
\usepackage{xcolor}
\usepackage{url}
\usepackage{authblk}
\usepackage[inline]{enumitem}
\def\BibTeX{{\rm B\kern-.05em{\sc i\kern-.025em b}\kern-.08em
		T\kern-.1667em\lower.7ex\hbox{E}\kern-.125emX}}

\begin{document}
	\title{The TeamPlay Project: \\ Analysing and Optimising Time, Energy, and Security for Cyber-Physical Systems
		\thanks{$^{*}$ During the execution of the TeamPlay
                  project B.~Rouxel was at~$^\text{10}$, Y.~Marquer
                  at~$^\text{1}$, and A.~Seewald at~$^\text{9}$.
                $^\text{**}$ The TeamPlay project was coordinated by
                O.~Zendra $<$Olivier.Zendra@inria.fr$>$.}
	}

\author[1]{Benjamin Rouxel$^{*}$}
\author[2]{Christopher Brown}
\author[3]{Emad Ebeid}
\author[4]{Kerstin Eder}
\author[5]{Heiko Falk}
\author[6]{Clemens Grelck}
\author[7]{\\Jesper Holst}
\author[5]{Shashank Jadhav}
\author[8]{Yoann Marquer$^{*}$}
\author[9]{Marcos Martinez De Alejandro}
\author[4]{Kris Nikov}
\author[3]{Ali Sahafi}
\author[3]{\\Ulrik Pagh Schultz Lundquist}
\author[10]{Adam Seewald$^{*}$}
\author[11]{Vangelis Vassalos}
\author[12]{Simon Wegener}
\author[13]{Olivier Zendra$^{**}$}

\affil[1]{\textit{Unimore}, Italy}
\affil[2]{\textit{University of St.Andrews}, United Kingdom}
\affil[3]{\textit{University of Southern Denmark}, Denmark}
\affil[4]{\textit{University of Bristol}, United Kingdom}
\affil[5]{\textit{Hamburg University of Technology}, Germany}
\affil[6]{\textit{University of Amsterdam}, Netherlands}
\affil[7]{\textit{SkyWatch A/S}, Denmark}
\affil[8]{\textit{University of Luxembourg}, Luxembourg}
\affil[9]{\textit{Thales Alenia Space}, Spain}
\affil[10]{\textit{Yale University}, USA}
\affil[11]{\textit{Irida Labs AE}, Greece}
\affil[12]{\textit{AbsInt Angewandte Informatik GmbH}, Germany}
\affil[13]{\textit{INRIA, University of Rennes, CNRS, IRISA}, France}

\setcounter{Maxaffil}{0}
\renewcommand\Authands{, }

	\maketitle

	\begin{abstract}
				Non-functional properties, such as \emph{energy}, \emph{time}, and \emph{security} (ETS) are becoming increasingly
		important in Cyber-Physical Systems (CPS) programming.
		This article describes TeamPlay, a research project funded under the \textbf{EU Horizon 2020 programme} between January
		2018 and June 2021. 
		TeamPlay aimed to provide the system designer with a toolchain for developing embedded applications where ETS properties are first-class citizens, allowing the developer to reflect directly on energy, time and security properties at the source code level.
	%
%
%
		In this paper we give an overview of the TeamPlay methodology, introduce the challenges and solutions of our approach and summarise
		the results achieved. 
		Overall, applying our TeamPlay methodology led to an improvement of up to 18\% performance and 52\% energy usage over traditional approaches. 
	\end{abstract}


	\section{Introduction}
%
Non-functional properties, such as \emph{energy}, \emph{time}, and \emph{security} (ETS), are increasingly important for the programming of cyber-physical systems (CPS). 
These systems typically include small embedded devices, which operate under very restrictive ETS constraints, and rely extensively on battery power. 
Optimising both application execution time and energy consumption while meeting the specified security requirements must be performed early during development to achieve the tight time-to-market windows in this sector with high confidence into the correctness of the CPS.
Developing embedded software that respects its specified ETS budget is challenging for developers, who frequently lack tool support to expose and analyse ETS properties during development. 
Traditional techniques rely on dynamic profiling of code on the embedded device, followed by a time-consuming iterative process of understanding the profiling data and several trial-and-error optimisation attempts. 
Furthermore, developing software that is optimised for multiple non-functional properties at once is an even greater challenge.
The TeamPlay project addressed these challenges by providing the developer with tool support to analyse ETS properties at the source code level together with a methodology for developing embedded applications in which ETS properties are treated as \emph{first-class citizens}.
The TeamPlay approach, therefore, enables developers to reason about the energy, time, and security properties of their applications very early in the process and as an integral part of their typical development activities.
The Teamplay methodology comprises a number of key components:

\begin{itemize}
\item A \textit{cross-layer management of ETS properties} from source code to binary, through a complex set of optimisations arising from the different elements of the developer workflow, i.e., the compiler and the scheduler.
  
\item A \emph{contract specification language (CSL)}, allowing ETS properties to be reflected into source code as a first-class citizen, and allowing contracts over those ETS properties to be expressed in the source and proven formally using dependent types.

\item An explicit \emph{coordination layer} that takes care of scheduling and mapping decisions on heterogeneous multi-core architectures.
  
\item \textit{Comprehensive energy modeling methods} for characterising the energy consumption of commercial off-the-shelf (COTS) platforms, and predicting energy usage either statically during compilation or dynamically at runtime.

\item \textit{Demonstration of the Teamplay methodology} on a number of realistic use cases from the medical, aerospace, autonomous systems, and AI domains where significant improvements in both ETS and in usability have been achieved.
\end{itemize}



In this paper,
(1) we give an overview of the TeamPlay toolchain and methodology for both \emph{predictable} and \emph{complex} architectures;
(2) we present a high-level overview of the key technical challenges that TeamPlay addressed, including making ETS properties first-class citizens in a \emph{transparent} manner, enabling the programmer to reason about ETS constraints as part of their development process, energy-modelling targetting multiple different architecture types and generally software development with stringent ETS goals and requirements from the outset;
and (3) we demonstrate the TeamPlay methodology on a number of industrial use cases from the domains of medical devices, space, aerial vehicles, and deep learning. 




	\section{Overview of the TeamPlay Toolchain}
\label{sec:overview}

The TeamPlay approach is focussed around providing programmer-friendly abstractions that turn ETS properties into first-class citizens of the language they are embedded in. This allows the programmer to annotate source code with timing, energy, and security information, assign
them to structures in the code, and write assertions that can be proven about these properties. By
exposing ETS properties as first-class citizens and by making them easy to use and
understand for application programmers, we enable the programmer to manipulate,
and precisely reason about, ETS as normal program values, directly interacting
with analytical 
and optimisation frameworks, including the physical properties of the underlying hardware.

Depending on the criticality and performance constraints of the system, we consider different platforms  ranging from simple \emph{predictable} hardware platforms, e.g., the Nucleo STM32F091RC,
to more \emph{complex} and less predictable ones, e.g., the Apalis TK1. 
The applicability of some sub-components of the methodology, especially time-oriented ones, strongly depends on the platform choice. In order to enable the TeamPlay methodology on the widest possible range of hardware options we design two specialized workflows, one targeting predictable architectures and one targeting complex architectures as explained in the two subsequent sections.


\subsection{The TeamPlay Methodology for Predictable Architectures}
\label{ssec:pred-arch}

We classify an architecture as \emph{predictable} if the number of cycles that an instruction takes to execute can be statically determined.
The deterministic nature of such processors makes them perfect targets to have their energy consumption modelled at the Instruction Set Architecture (ISA)
level, which in turn facilitates analysing security properties by observing time and energy consumption.

\begin{figure}
	\centering
	\includegraphics[width=\linewidth]{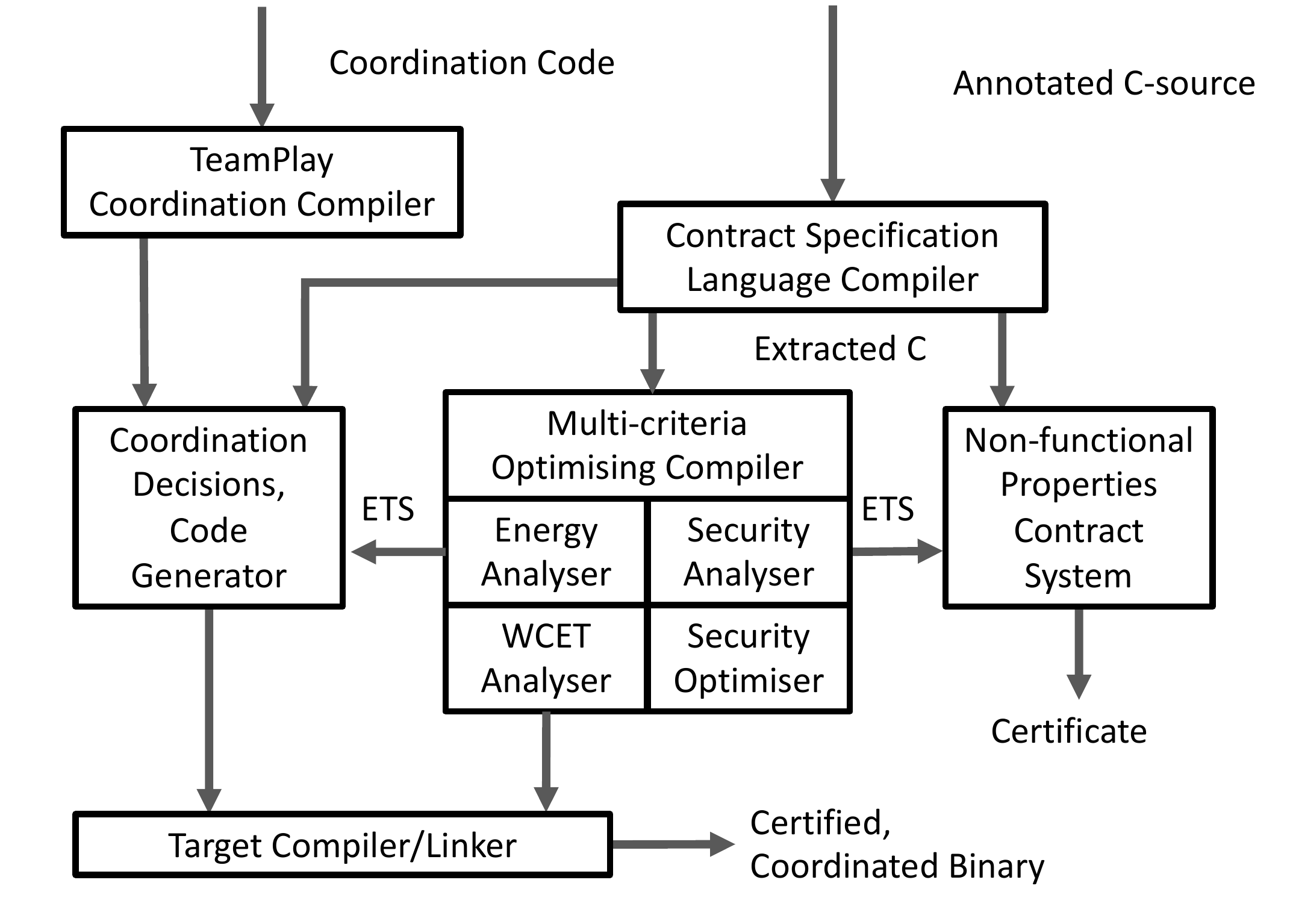}
	\caption{TeamPlay Methodology for Predictable Architectures.} 
	\label{fig:predict-arch}
\end{figure}

Figure~\ref{fig:predict-arch} depicts the workflow for such architectures. The methodology starts with an annotated version of the application source code in C. 
The annotations are defined in the Contract Specification Language (CSL)~\cite{D1.2}. 
They capture ETS properties at specific code locations.
The CSL layer gathers information about the code structure, e.g.,\ tasks and their parameters, and the associated ETS code points of interests (POIs).
The collected POIs are then sent to the multi-criteria optimising compiler (implemented in the WCET-aware C compiler WCC~\cite{Fal10,Roth:2018:SCOPES:3207719.3207729,Jadhav:2019:JRWRTC}), that applies a set of multi-objective optimisations~\cite{Jadhav:2019:SCOPES} regarding energy usage, timing control, and security risks.
To do this, WCC employs three plug-in tools, each of which targets a distinct non-functional property.
The aiT tool~\cite{ferdinand2004ait} statically computes a bound for the Worst-Case Execution Time (WCET) for each identified task in the code.
The EnergyAnalyser tool~\cite{D4.5} uses robust and accurate energy models~\cite{nikov2022robust,georgiou2021comprehensive} for the target hardware platforms to predict the energy consumed by each task.
The Security\-Analyser quantifies the security protection level against timing and power side-channel attacks with novel metrics~\cite{marquer:hal-03793085},
and the Security\-Optimiser transforms the code, if required, to increase protection against these side-channel attacks~\cite{MR20,BBM+22}.
%

In addition to generating object files for the input program, the multi-objective optimising compiler generates a file containing all the ETS properties extracted during compilation. 
Coupled with the information extracted from the source code, this enables the coordination layer~\cite{RoedRouxAltm+20} to validate the schedule and generate the necessary glue code for the initialisation, configuration, and runtime management of the tasks~\cite{RouxAltmGrel21}.

Similarly, the Non-functional Properties Contract System~\cite{brown2019type, 10.1145/3412932.3412944} formally proves, using dependent types, that both energy and time budgets as well as the security risk of each identified POI respects the ETS properties extracted by the compiler. The Contract System generates a certificate that could serve as a proof for certification authorities.

\subsection{The TeamPlay Methodology for Complex Architectures}
\label{ssec:unpred-arch}

In contrast to predictable architectures, \emph{complex} architectures cannot be statically analysed to determine WCETs due to features such as complex pipelines~\cite{heckmann2003influence} or generally undisclosed architectural and design information. 
Thus, the static analysers in the workflow for predictable architectures must be replaced by dynamic profilers for complex architectures.

\begin{figure}
	\centering
	\includegraphics[width=\linewidth]{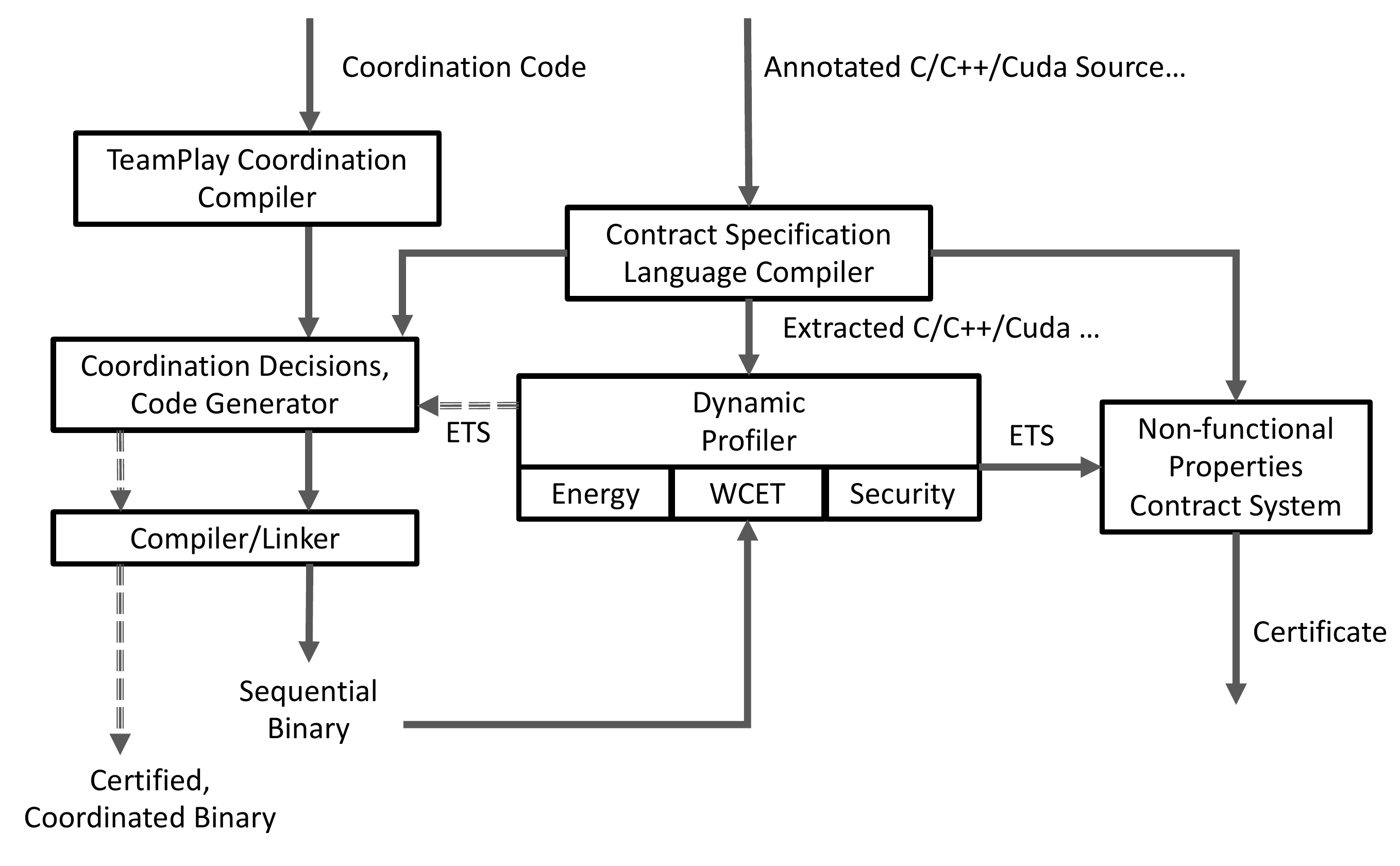}
	\caption{TeamPlay Methodology for Complex Architectures}
	\label{fig:complex-arch}
\end{figure}

Figure~\ref{fig:complex-arch} depicts the resulting workflow. 
In addition to C source code, other programming languages, including C++ and CUDA, can be annotated with ETS properties in CSL, demonstrating   the general applicability of CSL annotations.
The CSL layer identifies the POIs within the code and extracts the program structure, i.e., tasks and their dependencies. 
The coordination layer is then executed and, at first pass (solid arrow path), generates glue code to \textit{sequentially} manage the execution of tasks, while respecting dependencies. 
The resulting sequential binary is then instrumented with measuring points that, when executed several times, allow us to generate a dynamic profile of the energy consumption and execution time for each task. 
In this toolflow, energy measurement and execution time extraction are performed using PowProfiler~\cite{seewald2019component,seewald2019coarse}.
Reconnecting with the previous workflow, the CSL and coordination layer are provided with the estimated ETS properties of the application.
The coordination layer uses this information to schedule the application~\cite{roeder2021energy,RoedRouxGrel21}, generating glue code for the management of the tasks accounting for the parallel capability of the underlying platform (dashed path in Figure~\ref{fig:complex-arch}).

	\section{TeamPlay Challenges}
\label{sec:challenges}

\subsection{The Transparency Challenge}

During the design of a CPS it is common practice to analyse the ETS effects once the application has been completed.
The challenge we address with TeamPlay is to enable programmers to reason about these non-functional properties in an early stage of the development cycle instead.
TeamPlay promotes ETS information from the binary to the source code level.
Full transparency of ETS properties requires correlating source code features with the corresponding parts of the binary~\cite{entra}.  
A contract must be formally proven to guarantee the ETS constraints across the different levels of abstraction. Clear, human-understandable feedback needs to be provided in order to allow the developer to take actions should the application code fail to satisfy some of the constraints.

\subsection{The Energy Modelling Challenge}

A fundamental step to enable static energy consumption estimation of embedded software is the ability to accurately characterise and model the energy consumed by the target applications. This requires data collection from each target hardware platform and a model generation methodology.
The resulting energy models are directly sourced by static energy analysis tools.
Existing techniques may rely on available hardware support or external devices to associate costs with ISA-level constructs. 
However, these types of models often involve a very deep micro-architectural understanding of the target platform and can require a significant amount of time to develop and validate.
Since TeamPlay intends to target different types of architectures, a particular challenge is to design a configurable modelling methodology that captures platform behaviour at a higher abstraction level while yet providing very accurate prediction.
Additionally, the methodology must be cost-effective to promote wide-spread adoption. 

\subsection{The ETS-aware Development Challenge}

ETS properties are frequently considered as unrelated entities that can be accounted for independently of one another, despite the fact that there is substantial evidence that they are sometimes highly correlated. 
For some architectures, the longer an application executes, the larger the energy consumption and vice versa. However, this relationship is becoming less and less trivial in modern semiconductor systems since leakage power dissipation no longer decreases with reductions in transistor size, so there is often a "sweet-spot" somewhere in the middle of the frequency range where optimal energy consumption can be achieved. This is especially true for modern embedded systems.
Similar considerations can be made when security concerns are added, since some security countermeasures may imply executing additional instructions to obfuscate parts of the control flow that would lead to the leakage of sensitive information.
Considering all these non-functional properties at once is, therefore, the main challenge addressed in TeamPlay.

WCET analysis is used to determine how much time is spent in the worst case for the execution of a particular piece of software. It is thus a crucial part of the development process of safety-critical embedded systems.
Most existing WCET-aware compilation frameworks do not provide a smooth integration of WCET analysis into a compiler~\cite{KiPu01}, or do not scale with industrial-grade applications~\cite{ZKW+04a}.
Similar considerations apply to the integration of energy consumption and security analysis into compiler infrastructures.
In TeamPlay, we targetted full integration and exploitation of ETS analysers at the compilation level, making the use of these tools within each workflow transparent to the developer.

Worst-case optimisation demands fundamentally different optimisation strategies than average-case optimisation.
It is expected that, similar to time, average and worst-case energy usage will exhibit similar fundamental differences.
The overall challenge of ETS-aware compilation is thus to identify feasible optimisation approaches in order to balance energy usage, timing, and security guarantees within the compiler, both from the average and the worst-case perspective.
Unlike time (e.g. seconds or clock cycles) and energy (e.g. Joules), there is no consensus on one single objective security metric. Existing metrics are usually tied to particular attacks.
Consequently, we designed and implemented novel security metrics to counter timing and power side-channel attacks. 
Here, our focus is on information leakage related to time and energy/power.
Schedulers tend to follow the same path as compilers when dealing with energy, time, and security.
A challenge in TeamPlay was to raise the awareness of the scheduling strategy to embrace the complete set of ETS properties.

	\section{TeamPlay Use Cases and Key Results}
\label{sec:use-cases}

To validate the TeamPlay methodology, we apply it to four industrial-grade use cases.
The purpose of the use cases is to evaluate the benefits of the TeamPlay approach in terms of energy saving, timing control, and security improvement.
Some purely technical issues that could not be resolved within the project time-frame  hindered the evaluation of security on the industrial use cases; thus, our security approach and tools were validated on synthetic benchmarks on the Cortex-M0.
All evaluation results are provided in a public deliverable~\cite{d5.4}.

Through our evaluation, we observed that the TeamPlay methodology gives considerable reduction in development time and effort, and significantly increased confidence into the behaviour of applications regarding their ETS properties and requirements. However, we note that it is typically characteristic that such efficiency and confidence improvements cannot be easily  quantified objectively.

\subsection{Camera Pill}

For decades, Gastrointestinal (GI) cancer has been one of the leading causes of death in developed countries~\cite{Xie06}. Regular screening of high-risk patients with a genetic background can reduce mortality by detecting fatal diseases early.
Capsule Endoscopy enabled through camera pills has recently received a great deal of attention as a minimally invasive imaging procedure that offers novel methods for early diagnosis of disorders of the GI tract~\cite{Sahafi2022}.

A camera pill embeds an imaging system that takes thousands of pictures during its journey inside the GI tract, which are radio-transmitted to an external receiver.
The proper functioning of the camera pill is dependent on the timely execution of its software and hardware, as well as very stringent requirements in terms of power consumption, mostly due to its size.
Failure to meet timing constraints could result in a misleading or inconclusive diagnosis; failure to meet the energy budget could deplete the battery before reaching the area of interest. Furthermore, the transmitted images are medical data and hence subject to strict privacy regulations. This requires additional security measures to prevent interception.

Technically, the camera pill embeds a single-core Cortex-M0 coupled with a low-power FPGA acting as an image co-processor. During the lifetime of the project, we successfully applied the full 
TeamPlay toolchain for predictable architectures (Sec~\ref{ssec:pred-arch}), with the exception of the coordination layer, which could not be used due to a need for hardware-specific optimisations in the scheduler.
This processor architecture is fully supported by our static analysis tools, enabling safe and accurate time and energy estimation based on highly accurate energy models~\cite{georgiou2021comprehensive}. We evaluated security requirements and added encryption capabilities as needed. The WCC compiler successfully trades execution time with energy and generated highly optimised code. The coordination layer and CSL both give a green light with a valid schedulability analysis and a certificate proving the specified energy and time budgets being met.
%
Overall, applying the TeamPlay methodology \emph{led to an improvement of 18\% performance and 19\% energy usage over the use of traditional toolchains}.

\subsection{Communication in Space Applications}

Energy consumption is a major concern in spacecraft, where every little reduction of power consumption that can be achieved without a significant loss of performance is highly valuable. The space-oriented use case includes an image processing and transmission application using SpaceWire~\cite{parkes2005spacewire}.
The proper functioning of the full system is dependent on the timely execution of its software and hardware to avoid missing images. Moreover, the harsh environment requires the overall energy consumed to be minimal. Furthermore, the transmitted images can be of critical sensitivity and must be protected from malicious third-party access.

The target hardware device is the LEON3FT-based GR712RC development platform from Cobham-Gaisler~\cite{l3fgrainmodeldata}. The runtime environment is provided by a Real-Time Operating System (RTOS) named RTEMS (\url{https://www.rtems.org/}), which executes the image processing and communication program. Due to its predictable-by-design nature, the LEON3FT is fully supported by the TeamPlay toolchain for predictable architectures. The source code has been annotated for all three ETS properties.
A complete energy model has been built~\cite{nikov2022robust} and integrated into the EnergyAnalyser tool that is called by WCC for multi-criteria optimisation. The coordination layer and CSL both gave a green light with a valid schedulability analysis and a certificate proving that the code meets its specified energy and time budget. The Coordination layer also generated the required initialisation, management, and configuration code for the application within the RTEMS runtime ecosystem. Applying the TeamPlay methodology led to an \emph{energy improvement of 52\% while meeting all deadlines}.

\subsection{Uncrewed Aerial Vehicle}

Uncrewed Aerial Vehicles (UAVs) or drones are increasingly used in a variety of scenarios, enhancing the use of conventional robotic systems with the oversight of an agile and easy-to-deploy platform.
TeamPlay had two UAV use cases~\cite{EBEID201811}. The first concerns a search and rescue (SAR) UAV, a drone surveying above the sea seeking lifeboats. Upon detection, a ground base station is notified. The second use case aims at increasing the quality of agricultural produce, relying on diverse sensing mechanisms and facilitating the transition towards \textit{precision agriculture} (PA), by, e.g., detecting ground hazards. In both cases, fixed-wing drones, i.e., UAVs where
propellers provide thrust, wings lift, and maneuvers are performed utilizing control surfaces~\cite{seewald2022energy}, embed a computing payload connected to a camera~\cite{seewald2020mechanical}.
Object detection is performed on the payload, and any results are
transmitted to the ground station. 
The proper functioning of the 
use cases depend on
correct soft-timely execution and low energy consumption. Due
to overlapping frames, for instance, missing a few deadlines is possible
without missing objects, whereas energy constraints are tighter, directly impacting the flight time and thus coverage area~\cite{zamanakos2020energy}.

Three different hardware platforms have been tested: \begin{enumerate*}\item an Apalis TK1; \item Nvidia TX2; and, \item Nvidia Nano\end{enumerate*}. In all cases, Linux OS runs use case-specific software, e.g., C++/CUDA computer vision algorithms for detections on the GPU, and is eventually complemented with Robot Operating System (ROS) middleware~\cite{zamanakos2020energy,seewald2022energy}.
We 
applied 
the TeamPlay toolchain for unpredictable architectures 
but, as a result of platform complexity, 
omitted 
fact checker and SecurityAnalyser, which both 
require further investigation.

A 
study of the SAR use case, available in~\cite{rouxel2020prego}, exhibits 
an increment in 
timing control. 
Using our TeamPlay methodology, we observe an \emph{energy improvement of 18\%}, 
resulting in the flight time being increased by approximately 4 minutes. 
%
Conversely to the SAR use case, the PA use case~\cite{seewald2020mechanical,zamanakos2020energy,seewald2022energy} utilized merely energy analysis, yet enabled flight time optimisations. When cruising, the mechanical components of the UAV consumed 28 Watts on average, whereas software components consumed between 2 and 11 Watts, with the toolchain enabling in-flight battery-aware schedulability~\cite{seewald2022energy}.

\subsection{Deep Learning Deployment}

The deep learning (DL) use case involves the use of convolutional neural networks (CNNs) for carrying out various image recognition tasks. CNNs are mathematical structures whose implementation is computationally demanding, involving convolution operations
between large tensors or multiplication of matrices of large dimensions. 
The DL use case is a free spot car parking detection application. We assume a camera placed above ground with multiple parking spots on sight and the trained CNN must then determine how many free spots are present and report it to an external receiver.
The goal is to study the application of ETS-oriented design which can then be ported to more advanced image recognition systems, such as industrial automation, or the space sector.

Two different hardware platforms have been considered: \begin{enumerate*}\item a single-core Cortex-M0; and, \item an Nvidia TK1\end{enumerate*}.  Our results show that the multi-criteria optimising compiler offers different compiled variants of the same tasks with different energy consumptions and WCET characteristics that can be a great guide for the application designer.

The TK1 version used only the coordination layer of the unpredictable architectures toolchain with a manual extraction of the application structure and a custom energy and time estimation. Results show that the application generated from the TeamPlay toolchain performs \emph{similarly as the original human-optimized version both in terms of energy and time}. 

	\section{Conclusion}
\label{sec:conclusion}

This paper introduced and summarised \emph{TeamPlay}, a research project supported by the EU Horizon-2020 programme from January 2018 to June 2021.

The goal of TeamPlay was to provide the system designer with a functioning toolchain for developing embedded applications where energy, time and security properties (ETS) are first-class citizens. 
Our technologies enable developers to incorporate ETS as a design goal of the software, e.g., to meet time or energy budgets while satisfying security requirements.
We considered two architectural classes that we label \emph{predictable} and \emph{complex}. 
In this paper we provided an overview of the TeamPlay methodology for both architecture classes. 
We further demonstrated our methodology on industrial use cases from a variety of sectors, including medical, space, aerial vehicles, and deep learning, where, in some cases, we were able to achieve energy savings of up to 52\%.

In the future, we intend to expand our work in a number of directions. 
The TeamPlay methodology is currently restricted to C and C++ applications for embedded devices; we therefore plan to extend our methodology to cover a much broader range of general-purpose programming languages (such as, e.g., Java, Python or Haskell), as our proposed methodology is, in principle, generic, and the concepts could be ported to other systems. 
In addition, we intend to expand our energy-modelling techniques to a wider range of hardware types and to improve our methodology and tool support to design software with ETS properties as a major requirement. 
Preliminary work has shown that program transformation techniques, such as refactoring tool support, would be very applicable here and thus form a natural extension to our methodology.

%
%
%
%

\section*{Acknowledgements}
This work was supported by the EU Horizon-2020 project \emph{TeamPlay} (\url{https://www.teamplay-h2020.eu}), grant \#779882.

	\bibliographystyle{IEEEtran}
	\bibliography{biblio}

\begin{thebibliography}{10}
\providecommand{\url}[1]{#1}
\csname url@samestyle\endcsname
\providecommand{\newblock}{\relax}
\providecommand{\bibinfo}[2]{#2}
\providecommand{\BIBentrySTDinterwordspacing}{\spaceskip=0pt\relax}
\providecommand{\BIBentryALTinterwordstretchfactor}{4}
\providecommand{\BIBentryALTinterwordspacing}{\spaceskip=\fontdimen2\font plus
\BIBentryALTinterwordstretchfactor\fontdimen3\font minus
  \fontdimen4\font\relax}
\providecommand{\BIBforeignlanguage}[2]{{%
\expandafter\ifx\csname l@#1\endcsname\relax
\typeout{** WARNING: IEEEtran.bst: No hyphenation pattern has been}%
\typeout{** loaded for the language `#1'. Using the pattern for}%
\typeout{** the default language instead.}%
\else
\language=\csname l@#1\endcsname
\fi
#2}}
\providecommand{\BIBdecl}{\relax}
\BIBdecl

\bibitem{D1.2}
{TeamPlay Consortium}, ``{Deliverable D1.2: Report on Initial Implementation of
  Proof Library, including Initial Contract Specification Language
  Implementation},'' 2020.

\bibitem{Fal10}
H.~Falk and P.~Lokuciejewski, ``{A Compiler Framework for the Reduction of
  Worst-Case Execution Times},'' \emph{Real-Time Systems}, vol.~46, no.~2, pp.
  251--298, 2010.

\bibitem{Roth:2018:SCOPES:3207719.3207729}
\BIBentryALTinterwordspacing
M.~Roth, A.~Luppold, and H.~Falk, ``{Measuring and Modeling Energy Consumption
  of Embedded Systems for Optimizing Compilers},'' in \emph{21st International
  Workshop on Software and Compilers for Embedded Systems}, ser. SCOPES
  '18.\hskip 1em plus 0.5em minus 0.4em\relax New York, NY, USA: ACM, 2018, pp.
  86--89. [Online]. Available:
  \url{https://tore.tuhh.de/bitstream/11420/1754/1/20180529-scopes-roth.pdf}
\BIBentrySTDinterwordspacing

\bibitem{Jadhav:2019:JRWRTC}
\BIBentryALTinterwordspacing
S.~Jadhav, M.~Roth, H.~Falk, C.~Brown, and A.~Barwell, ``{Reasoning about
  non-functional properties using compiler intrinsic function annotations},''
  in \emph{13th Junior Researcher Workshop on Real-Time Computing}, ser. JRWRTC
  '19, 2019, pp. 25--28. [Online]. Available:
  \url{https://doi.org/10.15480/882.2545}
\BIBentrySTDinterwordspacing

\bibitem{Jadhav:2019:SCOPES}
\BIBentryALTinterwordspacing
S.~Jadhav and H.~Falk, ``{Multi-Objective Optimization for the Compiler of
  Real-Time Systems based on Flower Pollination Algorithm},'' in \emph{22nd
  International Workshop on Software and Compilers for Embedded Systems}, ser.
  SCOPES '19.\hskip 1em plus 0.5em minus 0.4em\relax New York, NY, USA: ACM,
  2019, pp. 45--48. [Online]. Available:
  \url{https://tore.tuhh.de/bitstream/11420/2724/1/201905-scopes-jadhav.pdf}
\BIBentrySTDinterwordspacing

\bibitem{ferdinand2004ait}
C.~Ferdinand and R.~Heckmann, ``{aiT}: Worst-case execution time prediction by
  static program analysis,'' in \emph{Building the Information Society}.\hskip
  1em plus 0.5em minus 0.4em\relax Springer, 2004, pp. 377--383.

\bibitem{D4.5}
{TeamPlay Consortium}, ``{Deliverable D4.5: Report on Energy Usage Analysis and
  on Prototype},'' 2020,
  \url{https://gitlab.inria.fr/TeamPlay_Public/TeamPlay_Public_Deliverables/-/blob/master/D4.5.pdf}.

\bibitem{nikov2022robust}
K.~Nikov, M.~Martinez, S.~Wegener, J.~Nunez-Yanez, Z.~Chamski, K.~Georgiou, and
  K.~Eder, ``Robust and accurate fine-grain power models for embedded systems
  with no on-chip pmu,'' \emph{IEEE Embedded Systems Letters}, 2022.

\bibitem{georgiou2021comprehensive}
K.~Georgiou, Z.~Chamski, K.~Nikov, and K.~Eder, ``A comprehensive and accurate
  energy model for arm's cortex-m0 processor,'' \emph{arXiv preprint
  arXiv:2104.01055}, 2021.

\bibitem{marquer:hal-03793085}
\BIBentryALTinterwordspacing
Y.~Marquer, O.~Zendra, and A.~Heuser, ``{The Indiscernibility Methodology:
  quantifying information leakage from side-channels with no prior
  knowledge},'' Sep. 2022, working paper or preprint. [Online]. Available:
  \url{https://hal.inria.fr/hal-03793085}
\BIBentrySTDinterwordspacing

\bibitem{MR20}
\BIBentryALTinterwordspacing
Y.~Marquer and T.~Richmond, ``{A Hole in the Ladder: Interleaved Variables in
  Iterative Conditional Branching},'' in \emph{27th IEEE Symposium on Computer
  Arithmetic, ARITH-2020}.\hskip 1em plus 0.5em minus 0.4em\relax IEEE Xplore,
  2020, pp. 56--63. [Online]. Available:
  \url{https://hal.archives-ouvertes.fr/hal-02889212v1}
\BIBentrySTDinterwordspacing

\bibitem{BBM+22}
\BIBentryALTinterwordspacing
C.~Brown, A.~D. Barwell, Y.~Marquer, O.~Zendra, T.~Richmond, and C.~Gu,
  ``Semi-automatic ladderisation: Improving code security through rewriting and
  dependent types,'' in \emph{2022 ACM SIGPLAN International Workshop on
  Partial Evaluation and Program Manipulation}, ser. PEPM 2022.\hskip 1em plus
  0.5em minus 0.4em\relax New York, NY, USA: ACM, 2022, p. 14–27. [Online].
  Available: \url{https://doi.org/10.1145/3498886.3502202}
\BIBentrySTDinterwordspacing

\bibitem{RoedRouxAltm+20}
\BIBentryALTinterwordspacing
J.~Roeder, B.~Rouxel, S.~Altmeyer, and C.~Grelck, ``Towards energy-, time- and
  security-aware multi-core coordination,'' in \emph{22nd International
  Conference on Coordination Models and Languages (COORDINATION 2020), Malta},
  ser. Lecture Notes in Computer Science, S.~Bliudze and L.~Bocchi, Eds., vol.
  12134.\hskip 1em plus 0.5em minus 0.4em\relax Springer, 2020, pp. 57--74.
  [Online]. Available:
  \url{https://www.lexuor.net/publications/Coordination2020.pdf}
\BIBentrySTDinterwordspacing

\bibitem{RouxAltmGrel21}
B.~Rouxel, S.~Altmeyer, and C.~Grelck, ``{YASMIN: a Real-time Middleware for
  COTS Heterogeneous Platforms},'' in \emph{22nd ACM/IFIP International
  Middleware Conference (MIDDLEWARE 2021)}.\hskip 1em plus 0.5em minus
  0.4em\relax ACM, 2021.

\bibitem{brown2019type}
\BIBentryALTinterwordspacing
C.~Brown, A.~D. Barwell, Y.~Marquer, C.~Minh, and O.~Zendra, ``Type-driven
  verification of non-functional properties,'' in \emph{21st International
  Symposium on Principles and Practice of Declarative Programming}, ser. PPDP
  '19.\hskip 1em plus 0.5em minus 0.4em\relax New York, NY, USA: ACM, 2019.
  [Online]. Available: \url{https://hal.inria.fr/hal-02314723/document}
\BIBentrySTDinterwordspacing

\bibitem{10.1145/3412932.3412944}
\BIBentryALTinterwordspacing
A.~D. Barwell and C.~Brown, ``{A Trustworthy Framework for Resource-Aware
  Embedded Programming},'' in \emph{31st Symposium on Implementation and
  Application of Functional Languages}, ser. IFL '19.\hskip 1em plus 0.5em
  minus 0.4em\relax New York, NY, USA: Association for Computing Machinery,
  2019. [Online]. Available: \url{https://doi.org/10.1145/3412932.3412944}
\BIBentrySTDinterwordspacing

\bibitem{heckmann2003influence}
R.~Heckmann, M.~Langenbach, S.~Thesing, and R.~Wilhelm, ``The influence of
  processor architecture on the design and the results of {WCET} tools,''
  \emph{Proceedings of the IEEE}, vol.~91, no.~7, pp. 1038--1054, 2003.

\bibitem{seewald2019component}
\BIBentryALTinterwordspacing
A.~Seewald, U.~P. Schultz, J.~Roeder, B.~Rouxel, and C.~Grelck,
  ``Component-based computation-energy modeling for embedded systems,'' in
  \emph{Proceedings Companion of the 2019 ACM SIGPLAN International Conference
  on Systems, Programming, Languages, and Applications: Software for
  Humanity}.\hskip 1em plus 0.5em minus 0.4em\relax ACM, 2019, pp. 5--6.
  [Online]. Available: \url{https://adamseewald.cc/short/component2019}
\BIBentrySTDinterwordspacing

\bibitem{seewald2019coarse}
\BIBentryALTinterwordspacing
A.~Seewald, U.~P. Schultz, E.~Ebeid, and H.~S. Midtiby, ``Coarse-grained
  computation-oriented energy modeling for heterogeneous parallel embedded
  systems,'' \emph{International Journal of Parallel Programming}, vol.~49,
  no.~2, pp. 136--157, 2019. [Online]. Available:
  \url{https://adamseewald.cc/short/coarse2019}
\BIBentrySTDinterwordspacing

\bibitem{roeder2021energy}
\BIBentryALTinterwordspacing
J.~Roeder, B.~Rouxel, S.~Altmeyer, and C.~Grelck, ``Energy-aware scheduling of
  multi-version tasks on heterogeneous real-time systems,'' in \emph{36th
  Annual ACM Symposium on Applied Computing}, ser. SAC'21.\hskip 1em plus 0.5em
  minus 0.4em\relax New York, NY, USA: ACM, 2021, p. 501–510. [Online].
  Available:
  \url{https://www.lexuor.net/publications/SAC_2021__Energy_aware_heterogeneous_scheduling.pdf}
\BIBentrySTDinterwordspacing

\bibitem{RoedRouxGrel21}
\BIBentryALTinterwordspacing
J.~Roeder, B.~Rouxel, and C.~Grelck, ``Scheduling {DAGs} of multi-version
  multi-phase tasks on heterogeneous real-time systems,'' in \emph{14th IEEE
  International Symposium on Embedded Multicore/Many-core Systems-on-Chip
  (MCSoC 2021), Singapore}.\hskip 1em plus 0.5em minus 0.4em\relax IEEE, 2021,
  to appear. [Online]. Available:
  \url{https://staff.fnwi.uva.nl/c.u.grelck/publications/2021_MCSoC21_multi_phase_scheduling.pdf}
\BIBentrySTDinterwordspacing

\bibitem{entra}
\BIBentryALTinterwordspacing
K.~Eder, J.~P. Gallagher, P.~López-García, H.~Muller, Z.~Banković,
  K.~Georgiou, R.~Haemmerlé, M.~V. Hermenegildo, B.~Kafle, S.~Kerrison,
  M.~Kirkeby, M.~Klemen, X.~Li, U.~Liqat, J.~Morse, M.~Rhiger, and
  M.~Rosendahl, ``Entra: Whole-systems energy transparency,''
  \emph{Microprocessors and Microsystems}, vol.~47, pp. 278--286, 2016.
  [Online]. Available:
  \url{https://www.sciencedirect.com/science/article/pii/S0141933116300862}
\BIBentrySTDinterwordspacing

\bibitem{KiPu01}
R.~Kirner and P.~Puschner, ``{Transformation of path information for WCET
  analysis during compilation},'' in \emph{13th Euromicro Conference on
  Real-Time Systems (ECRTS)}, Delft, Netherlands, 2001, pp. 29--36.

\bibitem{ZKW+04a}
W.~Zhao, P.~Kulkarni, D.~Whalley \emph{et~al.}, ``{Tuning the WCET of embedded
  applications},'' in \emph{9th IEEE Real-Time and Embedded Technology and
  Applications Symposium (RTAS)}, Toronto, Canada, 2004.

\bibitem{d5.4}
{TeamPlay Consortium}, ``{Deliverable D5.4: Report on Assessment of Project
  Outcomes},'' 2020.

\bibitem{Xie06}
X.~{Xie}, G.~{Li}, X.~{Chen}, X.~{Li}, and Z.~{Wang}, ``A low-power digital
  {IC} design inside the wireless endoscopic capsule,'' \emph{IEEE Journal of
  Solid-State Circuits}, vol.~41, no.~11, pp. 2390--2400, Nov 2006.

\bibitem{Sahafi2022}
\BIBentryALTinterwordspacing
A.~Sahafi, Y.~Wang, C.~L.~M. Rasmussen, P.~Bollen, G.~Baatrup, V.~Blanes-Vidal,
  J.~Herp, and E.~S. Nadimi, ``Edge artificial intelligence wireless video
  capsule endoscopy,'' \emph{Scientific Reports}, vol.~12, no.~1, p. 13723, Aug
  2022. [Online]. Available: \url{https://doi.org/10.1038/s41598-022-17502-7}
\BIBentrySTDinterwordspacing

\bibitem{parkes2005spacewire}
S.~Parkes and P.~Armbruster, ``Spacewire: a spacecraft onboard network for
  real-time communications,'' in \emph{14th IEEE-NPSS Real Time Conference,
  2005.}\hskip 1em plus 0.5em minus 0.4em\relax IEEE, 2005, pp. 6--10.

\bibitem{l3fgrainmodeldata}
\BIBentryALTinterwordspacing
K.~Nikov, M.~Martinez, P.~Vallejo, A.~Balbis, J.~Nunez-Yanez, and K.~Eder,
  ``{GR712RC} {LEON3} {P}ower {M}odel {D}ata,'' IEEE Dataport, 2021. [Online].
  Available: \url{https://dx.doi.org/10.21227/1y7r-am78}
\BIBentrySTDinterwordspacing

\bibitem{EBEID201811}
\BIBentryALTinterwordspacing
E.~Ebeid, M.~Skriver, K.~H. Terkildsen, K.~Jensen, and U.~P. Schultz, ``A
  survey of open-source {UAV} flight controllers and flight simulators,''
  \emph{Microprocessors and Microsystems}, vol.~61, pp. 11--20, 2018. [Online].
  Available:
  \url{https://www.researchgate.net/publication/325134452_A_Survey_of_Open-Source_UAV_Flight_Controllers_and_Flight_Simulators}
\BIBentrySTDinterwordspacing

\bibitem{seewald2022energy}
\BIBentryALTinterwordspacing
A.~Seewald, H.~Garc{\'i}a~de Marina, H.~S. Midtiby, and U.~P. Schultz,
  ``Energy-aware planning-scheduling for autonomous aerial robots,'' in
  \emph{Proceedings of the IEEE/RSJ International Conference on Intelligent
  Robots and Systems (IROS'22)}.\hskip 1em plus 0.5em minus 0.4em\relax IEEE,
  2022, pp. 2946--2953. [Online]. Available:
  \url{https://adamseewald.cc/short/energy2022}
\BIBentrySTDinterwordspacing

\bibitem{seewald2020mechanical}
\BIBentryALTinterwordspacing
A.~Seewald, H.~Garcia~de Marina, H.~S. Midtiby, and U.~P. Schultz, ``Mechanical
  and computational energy estimation of a fixed-wing drone,'' in \emph{2020
  Fourth IEEE International Conference on Robotic Computing (IRC)}.\hskip 1em
  plus 0.5em minus 0.4em\relax IEEE, 2020, pp. 135--142. [Online]. Available:
  \url{https://adamseewald.cc/short/mechanical2020}
\BIBentrySTDinterwordspacing

\bibitem{zamanakos2020energy}
\BIBentryALTinterwordspacing
G.~Zamanakos, A.~Seewald, H.~S. Midtiby, and U.~P. Schultz, ``Energy-aware
  design of vision-based autonomous tracking and landing of a {UAV},'' in
  \emph{2020 Fourth IEEE International Conference on Robotic Computing
  (IRC)}.\hskip 1em plus 0.5em minus 0.4em\relax IEEE, 2020, pp. 294--297.
  [Online]. Available: \url{https://adamseewald.cc/short/energy2020}
\BIBentrySTDinterwordspacing

\bibitem{rouxel2020prego}
\BIBentryALTinterwordspacing
B.~Rouxel, U.~P. Schultz, B.~Akesson, J.~Holst, O.~Jorgensen, and C.~Grelck,
  ``{PReGO}: a generative methodology for satisfying real-time requirements on
  cots-based systems: Definition and experience report,'' in \emph{19th ACM
  SIGPLAN International Conference on Generative Programming: Concepts and
  Experiences (GPCE 2020), Chicago, USA}.\hskip 1em plus 0.5em minus
  0.4em\relax ACM, 2020, pp. 70--83. [Online]. Available:
  \url{https://www.lexuor.net/publications/Drone_Use_Case_preprint.pdf}
\BIBentrySTDinterwordspacing

\end{thebibliography}
\end{document}